\documentclass[aps,prl,onecolumn,showpacs,groupedaddress,preprint]{revtex4}

\usepackage[T1]{fontenc}
\usepackage{epsfig}           
\usepackage{amsmath}

\begin{document}


\title{Size and Time Rescaling at the Paraelectric to Ferroelectric Phase Transition in BaTiO$_{\text{3}}$}

\author{Marek Pa\'{s}ciak$^1$, Salah Eddine Boulfelfel$^2$, and Stefano Leoni$^2$}
\affiliation{$^1$Institute of Low Temperature and Structure Research, Polish Academy of Sciences, P.O. Box 1410, 50-950 Wroc\l{}aw, Poland\\$^2$Max-Planck-Institut f\"ur Chemische Physik fester Stoffe, N\"othnitzer Str. 40, D-01187 Dresden, Germany}

\pacs{77.80.-e, 02.70.Ns, 07.05.Tp, 77.84.Dy}

\begin{abstract}
We describe length and time rescaling across the paraelectric to ferroelectric 
phase transition in BaTiO$_3$. Small ferroelectric clusters are unable to grow in the paraelectric phase and exist for a short period of time only. The onset of the phase transition invokes a rescaling of both size and time lenghts as necessary ingredients for ferroelectricity to come into existence. The growing of ferroelectric domains involves correlated chains of Ti-O dipoles, which causes domain wall rigidity and domain shape stability. Ferroelectric nanodomains may host islands of opposite polarization (antiferroelectric defects), which appear as an intrinsic feature of the growing BaTiO$_3$ ferroelectric material.
\end{abstract}

\maketitle

Ferroelectric materials are broadly employed as capacitors and memory
materials. However, even for the classical compound BaTiO$_3$ (BTO),
microscopic details on the cubic paraelectric (PE) and on the ferroelectric
(FE) phases and phase transformations (from cubic to tetragonal, orthorhombic
and rhombohedral phases on lowering temperature)
remains puzzling.  A macroscopic electrical polarization reflects some amount of coherent behavior in the atomic displacements. In BaTiO$_3$, the off-center diplacement of Ti atoms creates a net electric dipole moment. In terms of domain growth, a critical size \cite{spaldin} may thus be required for the ferroelectric state to be stable with respect to the competing electrostatic energy, caused by an asymmetric charge distribution. Recently, theoretical \cite{marco} and experimental \cite{fong} evidence on fundamental size limits in ferroelectricity has been provided, and important insights into device scalability have been collected. Simulations \cite{stachiotti, krakauer,krakauer-2} and experiments \cite{ziebinska,tai} indicate the existence of dynamic polar clusters already in the paraelectric, cubic phase, as local precursors of longer range ferrodomains. This suggests a \textit{rescaling of sizes} as a fundamental ingredient of ferroelectricity, rather than a size scale within a certain ferroelectric state. Rescaling was shown to play a key role in relaxor systems, recently \cite{dmowski}.
 
In this Letter, along this line, we perform MD simulations on the phase transition (PT) from PE cubic
(\textit{c}) to FE tetragonal (\textit{t}) in BTO ($T_c$= 393 K) for understanding ferroelectric domain formation. For this, we use transition path sampling molecular dynamics (TPSMD)\cite{dellago} as a means to elucidate mechanistic details of phase transitions \cite{dellago,prl-1,prl-2,chemistry, cdse} and to extract scale and time lengths (this work).

Two main models for ferroelectricity, 
the displacive \cite{cochran-displ} and the order-disorder model \cite{comes-od-1, bersunker-od-2, chaves-od-3} have been proposed over the years, which account for a subset of experimental facts \cite{lines}. 
In the displacive model, 
the softening of TO phonon modes detected
by neutron scattering \cite{harada} is microscopically connected
to Ti leaving the center of the oxygen octahedra at the transition
onset. EXAFS, XANES \cite{ravel} or X-ray diffuse scattering
highlight the relevance of <111> Ti displacements already in the cubic phase.
Molecular dynamics (MD) simulations \cite{stachiotti-2} indicate a crossover from displacive
soft mode to order-disorder dynamics close to the transition temperature (KNbO$_3$).
The implicit time averaging \cite{stern-nmr}
of NMR experiments \cite{zalar} leads to Ti distortion
along <100>, whereby the instantaneous displacements are along <111>
\cite{ravel}. The interaction of off-center Ti displacements and
soft modes are the main ingredients of the Grishberg and Yacoby model
\cite{girshberg-yacoby}, applied to BTO \cite{pirc}. 
Shell model \cite{stachiotti} and effective Hamiltonian MD
\cite{krakauer} point out the prevalent role of chain-like correlations, which
are not disappearing above $T_c$.
First principles calculations conclude the off-center position of Ti in cubic BTO \cite{cohen}. 
In a recent DFT-based approach \cite{goddard} the fundamental role of antiferroelectric (AFE)
coupling between Ti-O chains of opposite polarization is indicated. 

\begin{figure*}[t]
\epsfig{file=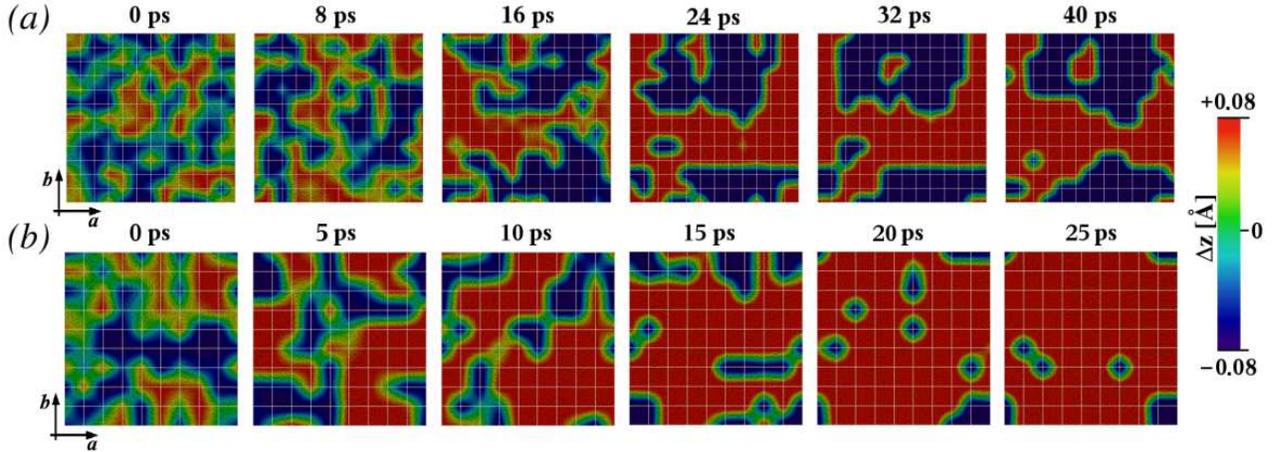,width=17.cm}
\caption{Transition path for temperature (a) and field induced (b) PE to FE phase transition. 
Ti displacements $\Delta z$ are shown in red and blue for (up) and
(down) domain polarization, respectively.
(a) 12$^{\text{3}}$ unit cells box. 
Clustering is clearly visible in the cubic phase (0-8 ps), just before domain growth onset (8-16 ps). 
(b) Under a static electric field $\vec{\cal E}$ the transformation rate is enhanced,
however certain chains remain antiparallel for a relatively long time ($\ge$ 25 ps). 
The box consists of 9$^{\text{3}}$ unit cells.}
\end{figure*}

The inclusion of local <111> Ti distortions, coupled to O shifts into
short-ranged chains is thus a key move for explaining
controversial aspects like active Raman signals in the cubic phases,
x-ray absorption fine structures, phonon modes damping and diffuse scattering.

While these facts strongly support the
central role of domains for ferroelectric properties, no clear
picture of domain formation across the ferroelectric transition has
been provided yet. Especially, the role and change of AFE and FE couplings,
and the spatial outreach of chains in different ferroelectric phases
remain unclear. 

To address these questions, we have performed TPSMD simulations on the paraelectric to ferroelectric phase transition in BaTiO$_3$, using the polarizable shell interatomic potential
of Sepliarsky \textit{et al.} \cite{sepliarsky} in the anisotropic NpT ensemble \cite{dlp}.

TPSMD is an iterative process and requires an initial trajectory \cite{dellago}. 
The latter was chosen in a regime of Ti positions shifted off-center from the ideal cubic arrangement, corresponding to the displacive model. This regime is quickly abandoned in the course of simulations, and a clear tendency to grow Ti-O chains along the {\it c}$\rightarrow${\it t} transition appeared. In the course of TPSMD a too narrow box would translate into an increased difficulty to stabilize the tetragonal configuration on chain growth. We used this \textit{intrinsic length rescaling} indicated by the simulation itself to define an appropriate size of the simulation system. Successive enlargement of the box dimensions showed that a 12$^{\text{3}}$ unit cells system is an adequate size to study ferrodomain formation.

The distinction between cubic and tetragonal structures was ensured by an order parameter,
based on monitoring cell parameter changes, and on collecting time-averaged
statistics over Ti displacements (see details below).

To better cover the diverse (microscopic \& macroscopic) aspects of the PE$\rightarrow$FE phase transition in BTO, 
we present in the following a threefold analysis of the simulation results:
(a) evaluation of instantaneous and time-averaged Ti-displacements
(Fig. 2) with respect to <111> directions (Fig. 4), for direct comparison with
EXAFS measurements (time-resolved method), (b) simulation of diffuse scattering (DS, Fig. 2) (c) elucidation of ferrodomain growth, domain morphologies and boundaries
(Figs. 1 \&3),

\begin{figure*}[t]
\epsfig{file=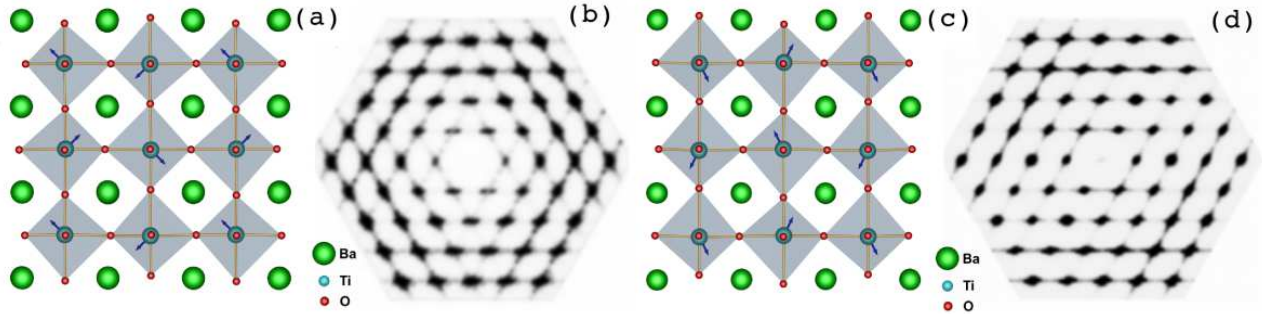,width=17cm}
\caption{Snapshots of the dynamic structure of BTO in (a) cubic PE phase and (c)
tetragonal FE phase. The arrows show the relative displacements of Ti
atoms from the centers of the oxygen octahedra. Note that Ba atoms
globally keep their positions and Ti atoms stay on the average in
the center of a Ba cube. The diffraction patterns on the \{111\}{*} reciprocal
plane contain diffuse scattering in the (b) cubic phase and (d) tetragonal
phase (Bragg scattering was subtracted). The set of \{001\}{*} diffuse
planes is clearly extincted in the tetragonal phase.}

\end{figure*}

\paragraph{Instantaneous and time-averaged Ti displacements.} Both PE and FE BTO phases host Ti displacements, connected over oxygen
into dynamic -(-Ti-O-)$_{\textit{m}}$- chains. The Ti environment is essentially rhombohedral
in the FE phase, due to the contraction of three Ti-O bonds and elongation
of the remaining three (cf. Fig. 2b), a pattern which is locally present
already in PE BTO. The degree of deviation of Ti displacements from
the closest <111> direction amounts 12.5 $^{\circ}$, in agreement with XAFS results, 11.8 $\pm$ 1.1$^{\circ}$, and comparable with the 10.8$^{\circ}$ based on the AFE/FE model \cite{goddard}.
Instantaneous projections of Ti displacements indicate <111> shifts
in both the PE and FE phases, whereby on time averaging they only
survive in the \textit{t} phase, along {[}001] (Fig. 4).

\paragraph{Diffuse scattering.} The existence of dynamic chains
gives rise to the appearance of X-ray DS in the form
of \{100\}{*}-type reciprocal planes intersecting \{111\}{*} plane
(Fig. 3) \cite{discus}. The extinction of a set of
directions (Fig. 2(d)) is due to Ti shift ordering in the FE phase.
The existence of locally anti-parallel arrangements of Ti shifts highlights
the role of AFE coupling within both FE and PE BTO.
The coexistence of FE and AFE couplings both in the
FE and the PE phases carries the appearance of DS. 
A modified pattern due to a changed chain timescale and shift ordering in the FE phase is responsible
for the extinction of a set of \{010\}{*} diffuse planes, an effect that can be suppressed upon doping \cite{Liu}.
The nature of DS is clearly related in PE and FE BTO. 
To better understand this point, especially with respect to the extent
of FE and AFE couplings, we now proceed to a full domain analysis
across the PT. 

\paragraph{Chains and ferrodomain growth.} To trace the progress of phase transition during TPSMD, besides cell
shape changes, we calculated transverse correlation \cite{transverse} and time autocorrelation of Ti-O
chain formations. We considered time-averaged Ti displacements over
200 fs. Transverse correlation is an appropriate quantity to monitor the progress of domain
growth as well as FE and AFE couplings, which appear as vector field
features in Fig. 1,a. Therein, different colors are associated
with different Ti displacement directions ($\Delta z$) and overall domain polarization,
red for upwards ($+$) and blue for downwards ($-$). The existence of polar
clusters already in the PE phase (0-8 ps) can be clearly seen, associated
with AFE/FE interactions. The (fast) time scale of cluster aggregation/decomposition
is less than 1 ps, for a cluster size up to 2-3 unit cells of locally
rhombohedral pattern, supporting recent experiments \cite{ziebinska,tai}.
However, setting in at $t\ge$8 ps, a dramatic change in chain dynamics
marks the onset of the transition. The dynamically correlated clusters
grow larger and literally freeze into domains. The overall slow down
of the process continues (20-30 ps) until a smooth landscape of 180$^{\circ}$
polar domains is reached ($t$$\ge$40 ps). A quantitative description of
the transition in terms of order parameter changes is given in Fig.
3. Relative variation of lattice parameters amounts
to $c/a$=1.01, in agreement with the experimental value of 1.009 
\cite{kwei-av_str}. 
The value of spontaneous polarization (Fig. 3) due to the formation of
domains is (on the average) vanishing, with a hidden tendency of the polarization $P_{z}$ component
to drift. The non-zero value of the transverse correlation (Fig. 3)
along all three Cartesian axes already in the PE phase is related
to polar cluster formation. The increase of transverse correlation,
which is a direct indicator of domain growth, surpasses the timescale
of the abrupt change of dynamics in the system, the latter measured
in terms of time autocorrelation (Fig. 3). The critical slowing down
and regime change takes place between 8-12 ps. At the crossover the
system transforms from floppy-dynamic (polar clusters) into almost
static (domains). During domain formation the \textit{z} component
of the velocity vectors $v_{z}$ of Ti positions decouples from the
other two and vanishes. In this regime, contrary
to the PE phase, any further domain evolution in the FE phase requires
a whole chain to flip. Clearly, {\it there is a marked rescaling of critical
lengths in going from PE to FE}, already noticed above, in terms of
size of polar regions, from clusters to domains, as well as with respect
to range and selection of AFE couplings.
A precise size rescaling of chain growth correlation 
lengths and a detailed
representation of AFE/FE couplings in the PE and FE phases, as well as domain morphologies are the main results of this analysis. 
Additionally our dynamic simulations provide details on the time scale of domain
formation and domain boundaries (Fig.1, green). In fact, while the PE phase is floppy and may host distant,
time-decorrelated polar clusters (coarse and mobile domain boundaries), 
FE domain formation is a slower
process, with long-ranged correlation lengths within domains and a
changed chain flip response (narrow and rigid domain walls). The latter is responsible for trapping
narrow AFE islands within larger domains (Fig. 1(a), 24, 32, 40 ps).
This effect is characteristic of the PT, and reflects an intrinsic
properties of BTO to accumulate AFE defects within FE domains. 
This is expected to affect the switching response of BTO.

\paragraph{Chains and ferrodomain growth, $\vec{\cal E}>0$.} To verify this 
effect we have performed another set of TPSMD simulations
under an external static electric field $\vec{\cal E}$ along $z$. This was achieved by applying different
initial values to the two transition branches (very large value for
\textit{c}$\rightarrow$\textit{t}, tiny for \textit{t}$\rightarrow$\textit{c}), and incorporating field value reduction (\textit{c}$\rightarrow$\textit{t})
or enhancement (\textit{t}$\rightarrow$\textit{c}) into TPSMD, until an equal value
was reached, 60 $kV/cm$. As the field favorizes {\it t} BTO, a $9^3$ unit cells simulation box suffices here.
The changes in the order parameters are represented in
Fig. 3. To a shortened transition time corresponds a
smoother change of the lattice parameters. The $P_z$ component of the
polarization increases up to a value of about 17.5 $\mu C/cm^{2}$, close
to the experimental value of 17.0 $\mu C/cm^{2}$ \cite{kwei-av_str}. 
Both transverse correlation and time autocorrelation
of chain growth indicate a two step mechanism, at 5-15 ps and at 15-20
ps. The transverse correlation reveals further domain growth after
20 ps. A pure electrostrictive scenario would imply growth of one
domain at the expenses of the other after 10 ps under the effect of
the field. In BTO however, the tendency of placing AF chain defects
within large FE domains is kept, and survive in the mature,
single domain BTO after 25 ps, to the extent that the final steps
are affected by AF chain dynamics only.

\begin{figure}[t!]
\center
\epsfig{file=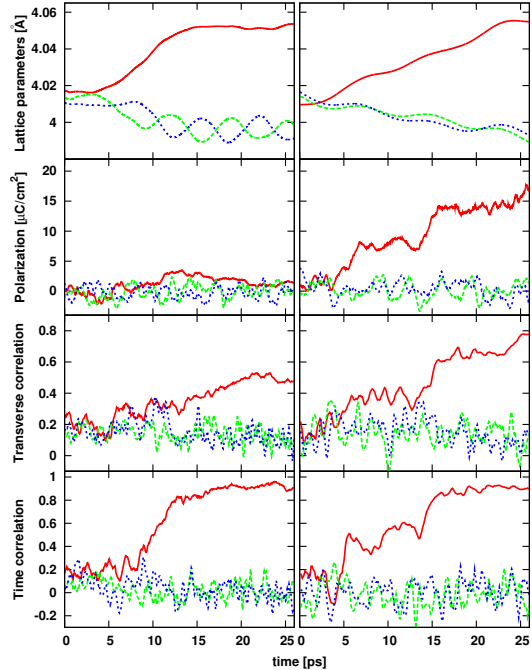, width=9.0cm}
\caption{Change of the structural characteristics monitored by different order
parameters. Left column corresponds to the tajectory of Fig. 1,a, right
column to Fig. 1,b. Colors: blue, green and red correspond to the $x$, $y$, and $z$ component, respectively. 
Time correlation is the chain auto-correlation (0.5 ps). 
Note that domain growth (transverse correlation \cite{transverse}) surpasses the timescale of chain mobility freezing.}
\end{figure}

\begin{figure}[t!]
\center
\epsfig{file=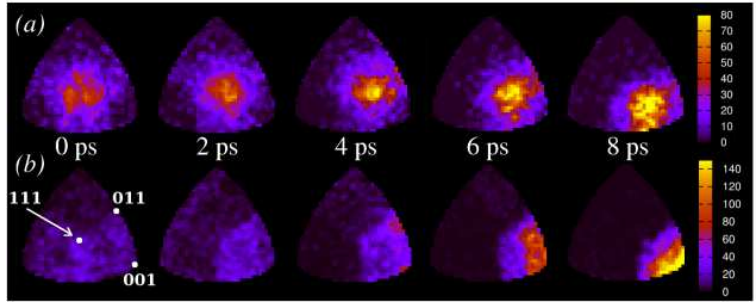,width=10.0cm}

\caption{Orthogonal projections of Ti shifts (first octant) during phase transition for (a)
0.05 ps and (b) 1 ps averaged structures. A clear
evidence that different timescales provide different pictures of the transition is obtained. }

\end{figure}

In conclusion, we have presented an atomistic investigation of the
paraelectric to ferroelectric phase transition in BTO. 
Dynamical polar subcritical clusters, 
already present in the cubic paraelectric phase, 
grow into larger ferroelectric nanodomains while the Ti-O dipole dynamics is slowing down.
Like in relaxors, local polar moments hints at localized phonons. Unlike relaxors, size rescaling in BTO restores a lattice dynamics more close to perfect ordered crystals (with the rhombohedral phase as limiting case of perfect dipole order), while in relaxors only the slow down process assists the formation of mesoscopic regions with local polarization. Along this line, the size \textit{and} time rescaling resulting from our TPSMD simulations may represent a general scheme for thinking about ordering phenomena in classic and relaxor ferroelectrics.\\
M.P. acknowledges the Polish Ministry of Education and Science (Grant \# N N202 023334) and the IMPRS in Dresden
(Klaus-Tschira Foundation scholarship). ZIH Dresden is acknowledged for computational time.


\begin{thebibliography}{24}

\bibitem{spaldin} N. A. Spaldin, Science \textbf{304}, 1606 (2004).

\bibitem{marco} M. Nu\~nez and M. Buongiorno Nardelli, Phys. Rev. Lett \textbf{101}, 107603 (2008).


\bibitem{fong} D. D. Fong \textit{et al.}, Science \textbf{304}, 1650 (2004).

\bibitem{stachiotti} S. Tinte {\it et al.},
Ferroelectrics \textbf{237}, 345 (2000).

\bibitem{krakauer} H. Krakauer {\it et al.},
J. Phys. Cond. Mat. \textbf{11}, 3779 (1999).

\bibitem{krakauer-2} R. Yu and H. Krakauer, Phys. Rev. Lett. \textbf{74}, 4067 (1995).

\bibitem{ziebinska} A. Zi\k{e}bi\'{n}ska {\it et al.},
J. Phys. Cond. Mat. \textbf{20}, 142202 (2008).

\bibitem{tai} R.Z. Tai {\it et al.},
Phys. Rev. Lett. \textbf{93}, 087601 (2004).

\bibitem{dmowski} W. Dmowski {\it et al.}, 
Phys. Rev. Lett. \textbf{100}, 137602 (2008).

\bibitem{dellago} C. Dellago {\it et al.}, 
Lect. Notes Phys. \textbf{703}, 349 (2006).

\bibitem{prl-1} D. Zahn and S. Leoni, 
Phys. Rev. Lett. \textbf{92}, 250201 (2004).

\bibitem{prl-2} S. E. Boulfelfel {\it et al.}, 
Phys. Rev. Lett. \textbf{99}, 125505 (2007).

\bibitem{chemistry} S. Leoni, 
Chem. Eur. J. \textbf{13}, 10022 (2007)

\bibitem{cdse} S. Leoni \textit{et al.}, 
Proc. Natl. Acad. Sci. \textbf{105}, 19612 (2008), 

\bibitem{cochran-displ} W. Cochran, 
Phys. Rev. Lett. \textbf{3}, 412 (1959).

\bibitem{comes-od-1} R. Comes {\it et al.},
Solid State Commun. \textbf{6}, 715 (1968).

\bibitem{bersunker-od-2} I. B. Bersuker, Phys. Lett. \textbf{20}, 589 (1966).

\bibitem{chaves-od-3} A. S. Chaves {\it et al.},
Phys. Rev. B \textbf{13}, 207 (1976).

\bibitem{lines} M.E. Lines, A.M. Glass, Principles and Applications of
Ferroelectrics and Related Materials (Clarendon, Oxford, 1977).

\bibitem{harada} J. Harada {\it et al.},
Phys. Rev. B \textbf{4}, 155 (1971).

\bibitem{ravel} B. Ravel {\it et al.},
Ferroelectrics \textbf{206}, 407 (1998).


\bibitem{stachiotti-2} M. Sepliarsky {\it et al.},
Comput. Mat. Sci. \textbf{10}, 51 (1998).


\bibitem{stern-nmr} E.A. Stern, Phys. Rev. Lett. \textbf{93}, 037601 (2004).

\bibitem{zalar} B. Zalar {\it et al.},
Phys. Rev. Lett. \textbf{90}, 037601 (2003).

\bibitem{girshberg-yacoby} Y. Girshberg and Y. Yacoby, Solid State Commun. \textbf{103}, 425 (1997).

\bibitem{pirc} R. Pirc and R. Blinc, Phys. Rev. B \textbf{70}, 134107 (2004).


\bibitem{cohen} R. E. Cohen, Nature \textbf{358}, 136 (1992).

\bibitem{goddard} Q. Zhang {\it et al.},
Proc. Natl. Acad. Sci. \textbf{103}, 14695 (2006).



\bibitem{sepliarsky} M. Sepliarsky {\it et al.},
Curr. Opin. Solid State Mater. Sci. \textbf{9}, 107 (2005).

\bibitem{dlp} W. Smith and T. J. Forester, J. Mol. Graphics 14 (1996)




\bibitem{discus} Diffuse scattering was calculated with DISCUS \cite{discus-ref} on averaging over 0.1 ps.

\bibitem{discus-ref} Th. Proffen and R.B. Neder, J. Appl. Cryst. \textbf{30}, 171 (1997).

\bibitem{Liu} Y. Liu {\it et al.},
Appl. Phys. Lett. \textbf{91}, 152907 (2007).

\bibitem{transverse} The transverse correlation of the $x$ ($y$) component of displacement is calculated as:
$\frac{1}{N}{\sum\limits_{i,j}} \sigma^{\xi}_{i}\cdot \sigma^{\xi}_{j}$, $\xi=x,y$.
$\vec\sigma$ is the vector of Ti displacement. 
The indices $i,j$ refer to  neighbor cells in the transverse directions. 

\bibitem{kwei-av_str} G. H. Kwei {\it et al.},
J. Phys. Chem. \textbf{97}, 2368 (1993).



\end{thebibliography}

\end{document}